\documentclass[journal]{IEEEtran}
\usepackage[pdftex]{graphicx}

\ifCLASSINFOpdf
\else
\fi
\usepackage{amsmath}
\usepackage{amssymb}
\usepackage{array}

\usepackage{algorithmic}
\usepackage{algorithm}

\usepackage{cite}

\usepackage{url}

\usepackage[usenames, dvipsnames]{color}



\begin{document}
%
\title{Convolutional Neural Network based Metal Artifact Reduction in X-ray Computed Tomography}
%
%
%

\author{Yanbo~Zhang,~\IEEEmembership{Senior~Member,~IEEE,}
        and~Hengyong~Yu*,~\IEEEmembership{Senior~Member,~IEEE}

\thanks{Copyright \copyright ~2017 IEEE. Personal use of this material is permitted. However, permission to use this material for any other purposes must be obtained from the IEEE by sending a request to pubs-permissions@ieee.org.}

\thanks{This work was supported in part by NIH/NIBIB U01 grant EB017140 and R21 grant EB019074. \emph{Asterisks indicate corresponding authors.}}
\thanks{Y. Zhang and H. Y. Yu* are with the Department of Electrical and Computer Engineering, University of Massachusetts Lowell, Lowell, MA 01854, USA. (e-mail: hengyong-yu@ieee.org).}} 


\maketitle

\begin{abstract}
In the presence of metal implants, metal artifacts are introduced to x-ray CT images. Although a large number of metal artifact reduction (MAR) methods have been proposed in the past decades, MAR is still one of the major problems in clinical x-ray CT. In this work, we develop a convolutional neural network (CNN) based open MAR framework, which fuses the information from the original and corrected images to suppress artifacts. The proposed approach consists two phases. In the CNN training phase, we build a database consisting of metal-free, metal-inserted and pre-corrected CT images, and image patches are extracted and used for CNN training. In the MAR phase, the uncorrected and pre-corrected images are used as the input of the trained CNN to generate a CNN image with reduced artifacts. To further reduce the remaining artifacts, water equivalent tissues in a CNN image are set to a uniform value to yield a CNN prior, whose forward projections are used to replace the metal-affected projections, followed by the FBP reconstruction. The effectiveness of the proposed method is validated on both simulated and real data.  Experimental results demonstrate the superior MAR capability of the proposed method to its competitors in terms of artifact suppression and preservation of anatomical structures in the vicinity of metal implants.
\end{abstract}

\begin{IEEEkeywords}
X-ray computed tomography (CT), metal artifacts, convolutional neural networks, deep learning
\end{IEEEkeywords}

%
\IEEEpeerreviewmaketitle

\section{Introduction}
%
%
\IEEEPARstart{P}{atients} are usually implanted with metals, such as dental fillings, hip prostheses, coiling, \emph{etc}. These highly attenuated metallic implants lead to severe beam hardening, photon starvation, scatter, and so on. This brings strong star-shape or streak artifacts to the reconstructed CT images \cite{DeMan1999}. Although a large number of metal artifact reduction (MAR) methods have been proposed during the past four decades, there is still no standard solution \cite{Lars2016,Jessie2015, Mouton2013}. Currently, how to reduce metal artifacts remains a challenging problem in the x-ray CT imaging field.

Metal artifact reduction algorithms can be generally classified into three groups: physical effects correction, interpolation in projection domain and iterative reconstruction. A direct way to reduce artifacts is to correct physical effects, \emph{e.g.}, beam hardening \cite{Park2016, Hsieh2000, Yanbo_BHC_2010} and photon starvation \cite{Marc2001}. However, in the presence of high-atom number metals, errors are so strong that the aforementioned corrections cannot achieve satisfactory results. Hence, the metal-affected projections are assumed as missing and replaced with surrogates \cite{Zaidi2013,zhang2011efficient,zhang2011new}. Linear interpolation (LI) is a widely used MAR method, where the missing data is approximated by the linear interpolation of its neighboring unaffected projections for each projection view. The LI usually introduces new artifacts and distorts structures near large metals \cite{Kalender1987}. By comparison, by employing \emph{a priori} information, the forward projection of a prior image is usually a more accurate surrogate for the missing data \cite{Bal2006, Meyer2010, Prell2009, Wang2013,zhang2014CMAR}. The normalized MAR (NMAR) is a state-of-the-art prior image based MAR method, which applies a thresholding based tissue classification on the uncorrected image or the LI corrected image to remove most of the artifacts and produce a prior image \cite{Meyer2010}. In some cases, artifacts are so strong that some image pixels are classified into wrong tissue types, leading to inaccurate prior images and unsatisfactory results. The last group of methods iteratively reconstruct images from the unaffected projections \cite{Wang1996, Wang1999, Xiaomeng2011, Yanbo_WTV_2010} or weighted/corrected projections \cite{Lemmens2009}. With proper regularizations, the artifacts are suppressed in the reconstructed results. However, due to the high complexity of various metal materials, sizes, positions, and so on, it is hard to achieve satisfactory results for all cases using a single MAR strategy. Therefore, several researchers combined two or three types of MAR techniques as hybrid methods \cite{Yanbo2013} \cite{HMAR2013}, fusing the merits of various MAR techniques. Hence, the hybrid strategy has a great potential to obtain more robust and outstanding performance by appropriately compromising a variety of MAR approaches. 

Recently, deep learning has achieved great successes in the image processing and pattern recognition field. For example, the convolutional neural network (CNN) has been applied to medical imaging for low dose CT reconstruction and artifacts reduction \cite{Wang2016,Jin_TIP_2017, Chen_TMI_2017, Wolterink_2017, Lars_SPIE_2017, Chen2017, Lars_fully3d_2017, du2017stacked, chen2018learn,liu2018low}. In particular, the concept of deep learning was introduced to metal artifact reduction for the first time in 2017 \cite{Lars_SPIE_2017,Lars_SPIE_image_2017,Lars_fully3d_2017,Yanbo_CNNMAR_2017,zhang2017cnnmar,Park_sinogram_2017,Park_BH_2017}. Park \textit{et al.} employed a U-Net to correct metal-induced beam hardening in the projection domain \cite{Park_sinogram_2017} and image domain \cite{Park_BH_2017}, respectively. Their simulation studies showed promising results over hip prostheses of titanium. However, the beam hardening correction based MAR methods have limited capability for artifact reduction in the presence of high-Z metal. Gjesteby \textit{et al.} developed a few deep learning based MAR methods that refine the performance of the state-of-the-art MAR method, NMAR, with deep learning in the projection domain \cite{Lars_SPIE_2017} and image domain \cite{Lars_fully3d_2017,Lars_SPIE_image_2017}, respectively. The CNN was used to help overcome residual errors from the NMAR. While their experiments demonstrated that CNN can improve the NMAR effectively, remaining artifacts are still considerable. Simultaneously, based on the CNN, we proposed a general open framework for MAR \cite{Yanbo_CNNMAR_2017,zhang2017cnnmar}. This paper is a comprehensive extension of our previous work \cite{Yanbo_CNNMAR_2017}. We adopt the CNN as an information fusion tool to produce a reduced-artifact image from some other methods corrected images. Specifically, before the MAR, we build a MAR database to generate training data for the CNN. For each clinical metal-free patient image, we simulate the metal artifacts and then obtain the corresponding corrected images by several representative MAR methods. Without loss of generality, we apply a beam hardening correction (BHC) method and the linear interpolation (LI) method in this study. Then, we train a CNN for MAR. The uncorrected, BHC and LI corrected images are stacked as a three-channel image, which is the input data of CNN and the corresponding metal-free image is the target, and a metal artifact reduction CNN is trained. In the MAR phase, the pre-corrected images are obtained using the BHC and LI methods, and these two images and the uncorrected image are put into the trained CNN to obtain the corrected CNN image. To further reduce the remaining artifacts, we incorporate the strategy of prior image based methods. Specifically, a tissue processing step is introduced to generate a prior from the CNN image, and the forward projection of the prior image is used to replace metal-affected projections. The advantages of the proposed method are threefold. First, we combine the corrected results from various MAR methods as the training data. In the end-to-end CNN training, the information from different correction methods is captured and the merits of these methods are fused, leading to a higher quality image. Second, the proposed method is an open framework, and all the MAR methods can be incorporated into this framework. Third, this method is data driven. It has a great potential to improve the CNN capability if we continue increasing the training data with more MAR methods. The source codes of our proposed method are open\footnote{https://github.com/yanbozhang007/CNN-MAR.git}.

The rest of the paper is organized as follows. Section II describes the creation of metal artifact database and the training of a convolutional neural network. Section III develops the CNN based MAR method. Section IV describes the experimental settings. Section V gives the experimental results and analyzes properties of the proposed method. Finally, Section VI discusses some relevant issues and concludes the paper.

\section{Training of the Convolutional Neural Network }

There are two phases to train a convolutional neural network for MAR. First, we generate metal-free, metal-inserted and MAR corrected CT images to create a database. Then, a CNN is constructed and the training data is collected from the established database and used to train the CNN.

\subsection{Establishing a Metal Artifact Database}

At first, we need to create a CT image database for CNN training. In this database, for each case, metal-free, metal-inserted, and MAR methods processed images are included.

\subsubsection{Generating Metal-free and Metal-inserted Images}

\begin{figure}
  \begin{center}
  \includegraphics[width=3.5in]{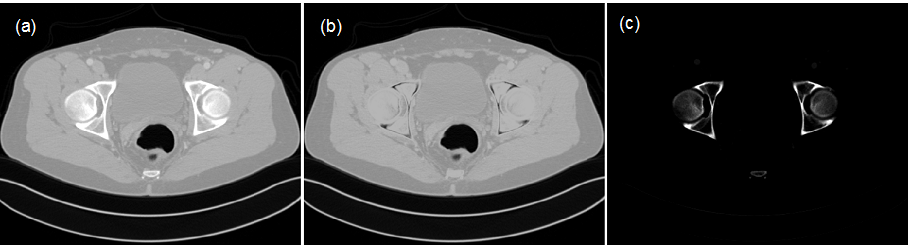}\\
  \caption{Example of tissue segmentation. (a) The benchmark image, (b) water-equivalent tissue and (c) bone.}
  \label{Fig1_segmentation}
  \end{center}
\end{figure}

In this subsection, we describe how to generate metal-free and metal-inserted CT images, where beam hardening and Poisson noise are simulated. To ensure that the trained CNN works for real cases, instead of using phantoms, we simulate the metal artifacts based on clinical CT images. To begin with, a number of DICOM format clinical CT images are collected from online resources and ``the 2016 Low-dose CT Grand Challenge'' training dataset \cite{AAPM2016}. In the presence of metal implants, we manually segment metals and store them as small binary images, which represent typical metallic shapes in real cases. Several representative metal-free CT images are selected as benchmark images. For a given benchmark image, its pixel values are converted from CT values to linear attenuation coefficients and denoted as $\mathbf{x}$. To simulate polychromatic projection, we need to know the material components in each pixel. Hence, a soft threshold-based weighting method \cite{EBHC2010} is applied to segment the image $\mathbf{x}$ into bone and water-equivalent tissue components, denoted as $\mathbf{x}^b$ and $\mathbf{x}^w$, respectively. Pixels with values below a certain threshold $T_1$ are viewed as water equivalent, while pixels above a higher threshold $T_2$ are assumed to be bone. The pixels with values between $T_1$ and $T_2$ are assumed to be a mixture of water and bone. Thus, a weighting function for bone is introduced as
\begin{equation}
\label{eq1}
\omega (x_i)=\left\{
	\begin{array}{ll}
		0, & x_i \le T_1\\
		1, & x_i \ge T_2\\
		\frac{x_i - T_1}{T_2 - T_1}, & T_1 < x_i <T_2
	\end{array}
\right. ,
\end{equation}
where $x_i$ is the $i^{th}$ pixel value of $\mathbf{x}$. Hence, $\mathbf{x}^b$ and $\mathbf{x}^w$ are expressed as 
\begin{equation}
\label{eq2}
x^b_i = \omega(x_i)x_i,
\end{equation}
\begin{equation}
\label{eq3}
x^w_i = (1 - \omega(x_i))x_i.
\end{equation}
Fig.~\ref{Fig1_segmentation} gives an example of the image segmentation.

For an x-ray path $L_j$, the linear integral of water and bone images are $d^w_j$ and $d^b_j$, respectively. We have
\begin{equation}
\label{eq4}
d^k_j = \int _{L_j} \mathbf{x}^k dl ,
\end{equation}
where the superscript ``$k$" indicates ``$w$'' or ``$b$''. To simulate polychromatic projections, we need to obtain linear attenuation maps of water and bone at various energies. For each material, the linear attenuation coefficient at the pixel is the product of the known energy-dependent mass attenuation coefficient and the unknown energy-independent density \cite{Elbakri2002}. We have
\begin{equation}
\label{eq5}
x^k_i(E) = m^k(E) \rho^k_i ,
\end{equation}
where $\rho^k_i$ is the density of ``$k$'' material at the $i^{th}$ pixel, and $m^w(E)$ and $m^b(E)$ are respectively mass attenuation coefficients at energy $E$ of water and bone. For a given polychromatic x-ray imaging system, let us assume that the equivalent monochromatic energy is $E_0$. Then, $x_i^b$ and $x_i^w$ can be written as
\begin{equation}
\label{eq6}
x^k_i = x^k_i(E_0) = m^k(E_0) \rho^k_i.
\end{equation}
Combining Eqs. (\ref{eq5}) and (\ref{eq6}), the unknown density $\rho^k_i$ can be eliminated. Hence, the energy dependent linear attenuation coefficient for each material is obtained as the following,
\begin{equation}
\label{eq7}
x^k_i(E) = \frac {x^k_i m^k(E)} {m^k(E_0)} .
\end{equation}
For the given x-ray path $L_j$, the ideal projection measurement $\bar y_j$ recorded by the $j^{th}$ detector bin is
\begin{equation}
\label{eq8}
\begin{array}{ll}
\bar y_j& = \int I(E) \text {exp} \left( - \int _{L_j} (x^w_i(E) + x^b_i(E)) dl \right) dE\\
    & = \int I(E) \text {exp} \left( - \int _{L_j} (\frac {x^w_i m^w(E)} {m^w(E_0)} + \frac {x^b_i m^b(E)} {m^b(E_0)}) dl \right) dE \\
    & = \int I(E) \text {exp} \left( - \frac {m^w(E) d^w_j} {m^w(E_0)} - \frac {m^b(E) d^b_j} {m^b(E_0)} \right) dE 
\end{array},
\end{equation}
where $I(E)$ is the known energy dependence of both the incident x-ray source spectrum and the detector sensitivity. Because the linear projection $d^w_j$ and $d^b_j$ have been computed in advance, computing the polychromatic projection using Eq. (\ref{eq8}) is very efficient. Approximately, the measured data follow the Poisson distribution:
\begin{equation}
\label{eq9}
y_j \sim \text{Poisson} \{ \bar y_j + r_j \},
\end{equation}
where $r_j$ is the mean number of background events and read-out noise variance, which is assumed as a known nonnegative constant \cite{Elbakri2002, Xu2012}. Thus, the noisy polychromatic projection $\mathbf{p}$ for reconstruction can be expressed as:
\begin{equation}
\label{eq10}
p_j = -\text{ln} \frac {y_i} {\int I(E) dE},
\end{equation}
The metal-free image is reconstructed using filtered backprojection (FBP), and the image is assumed as reference and denoted as $\mathbf{x}^{ref}$.

\begin{figure}
  \begin{center}
  \includegraphics[width=3.5in]{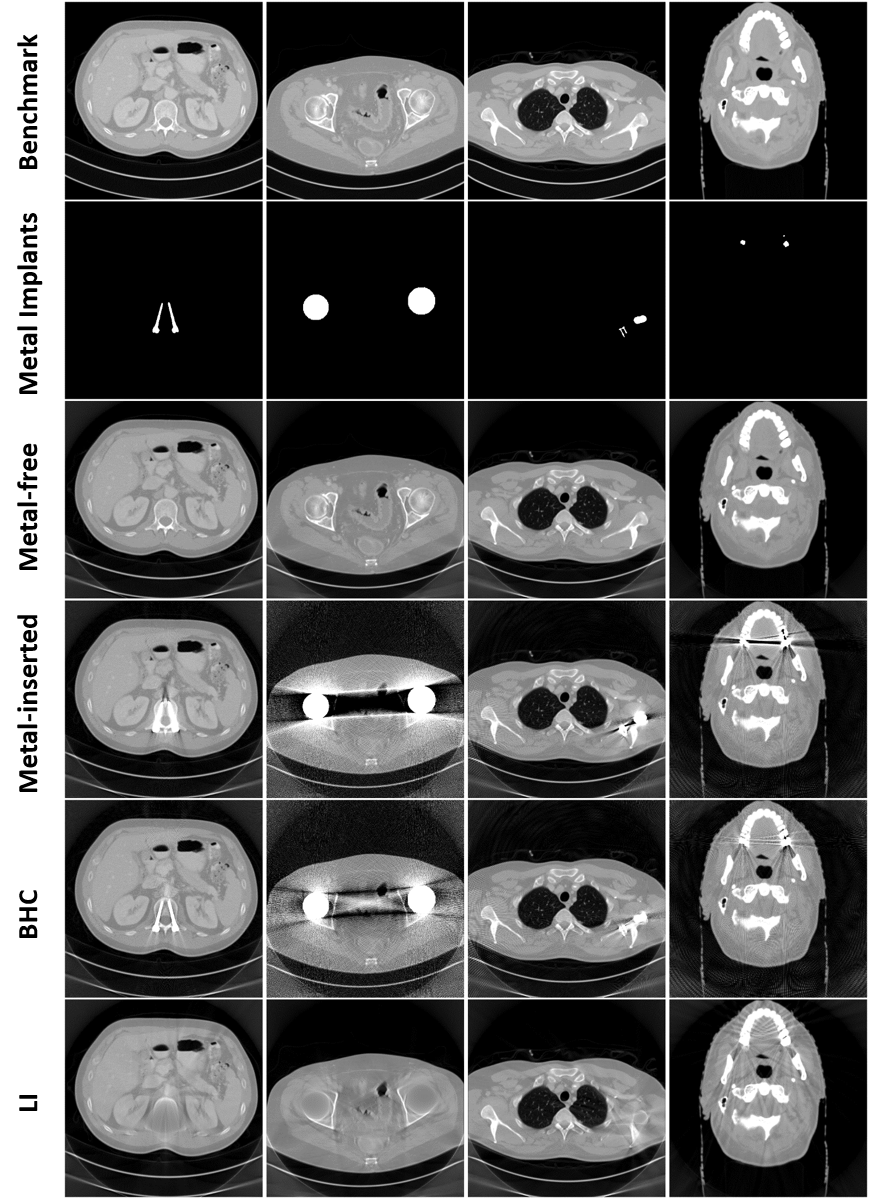}\\
  \caption{Representative samples in the database. Each column corresponds to one case. The top four rows are benchmark images, metal-only images, metal-free and metal-inserted images, respectively. The last two rows are images after metal artifact reduction using the BHC and LI, respectively.}
  \label{Fig2_imgDatabase}
  \end{center}
\end{figure}

To simulate metal artifacts, one or more binary metal shapes are placed into proper anatomical positions, generating a metal-only image $\mathbf{x}^m $. We specify the metal material, and assign metal pixels with linear attenuation coefficient of this material at energy $E_0$, and set the rest pixels to be zero. Because metals are inserted into patients, pixel values in $\mathbf{x}^b$ and $\mathbf{x}^w$ are set to be zero if the corresponding pixels in $\mathbf{x}^m$ are nonzero. Then, the $d^w_j$ and $d^b_j$ are updated, and the corresponding metal projection is computed using Eq. (\ref{eq4}). Similar to Eq. (\ref{eq8}), the ideal projection measurement is
\newcommand*{\Scale}[2][4]{\scalebox{#1}{$#2$}}%
\begin{equation}
\label{eq11}
\Scale[0.95]{
\bar y_j ^* = \int I(E) \text {exp} \left( - \frac {m^w(E) d^w_j} {m^w(E_0)} - \frac {m^b(E) d^b_j} {m^b(E_0)} - \frac {m^m(E) d^m_j} {m^m(E_0)} \right) dE , }
\end{equation}
where $m^m (E)$ is the mass attenuation coefficient of the
metal at energy $E$. Following the same operations in Eqs. (\ref{eq9})
and (\ref{eq10}), the noisy polychromatic projection $\mathbf{p}^*$ is obtained,
and then the image $\mathbf{x}^{art}$ containing artifacts is reconstructed.
Fig. \ref{Fig2_imgDatabase} shows four samples in the database. The top four rows
in Fig. \ref{Fig2_imgDatabase} are benchmark images, metal-only images, metal-free
images and metal-inserted images, respectively.

\subsubsection{Simple Metal Artifact Reduction}
We apply two simple metal artifact reduction methods, the linear interpolation (LI) and beam hardening correction (BHC) \cite{Verburg2012}, to alleviate artifacts. These methods are fast and easy to implement, and there are no manually selected parameters. Moreover, they suppress metal artifacts with different schemes, which have a great potential to provide complementary information for the CNN. In the LI method, the metal-affected projections are identified and replaced with the linear interpolation of their unaffected neighboring projections in each projection view. The LI corrected image is denoted as $\mathbf{x}^{LI} $. The BHC approach \cite{Verburg2012} adopts a first-order model of beam hardening error to compensate for the metal-affected projections. The length of metal $\{ l^m_j \}$ along each x-ray path is computed by forward projecting the binary metal-only image. The difference $\{ p^m_j \}$ between the original and LI projections is assumed as the contribution of metal. The correction curve between $\{ l^m_j \}$ and $\{ p^m_j \}$ is fitted to the correlation using a least squares cubic spline fit. Finally, the correction curve is subtracted from the original projection to yield the corrected data. The image obtained using BHC is denoted as $\mathbf{x}^{BHC}$. The two bottom rows in Fig. \ref{Fig2_imgDatabase} are four samples of BHC and LI corrected images, where metals are not inserted back into the LI images.

\subsection{Training a Convolutional Neural Network (CNN)}

For each sample in the database, the original uncorrected image, BHC image and LI image are combined as a three-channel image. The samples in the database are randomly divided into two groups for CNN training and validation. Small image patches of $s \times t \times 3$ are extracted from three-channel images, and these patches are assumed as the input data of CNN. Correspondingly, image patches of $s \times t$ are also obtained from the same positions of the metal-free images, and these patches are assumed as the target of CNN during training. The $r^{th}$ training sample pair is denoted as $\mathbf{u}_r \in \mathbb{R}^{s  \times t  \times 3 } $ and $\mathbf{v}_r \in \mathbb{R}^{s  \times t  } $, $r=1, \dots, R $, where $R$ is the number of training samples. The CNN training is to find a function $H: \mathbb{R}^{s  \times t  \times 3 } \to \mathbb{R}^{s  \times t }$ that minimizes the following cost function \cite{Chen2017}:
\begin{equation}
\label{eq12}
H = \text{arg} \min_H \frac {1}{R} \sum _{r = 1} ^R \| H(\mathbf{u}_r) - \mathbf{v}_r    \| _F ^2 , 
\end{equation}
where $ \| \cdot \| _F $ is the Frobenius norm.

\begin{figure*}
  \begin{center}
  \includegraphics[width = 0.95\textwidth]{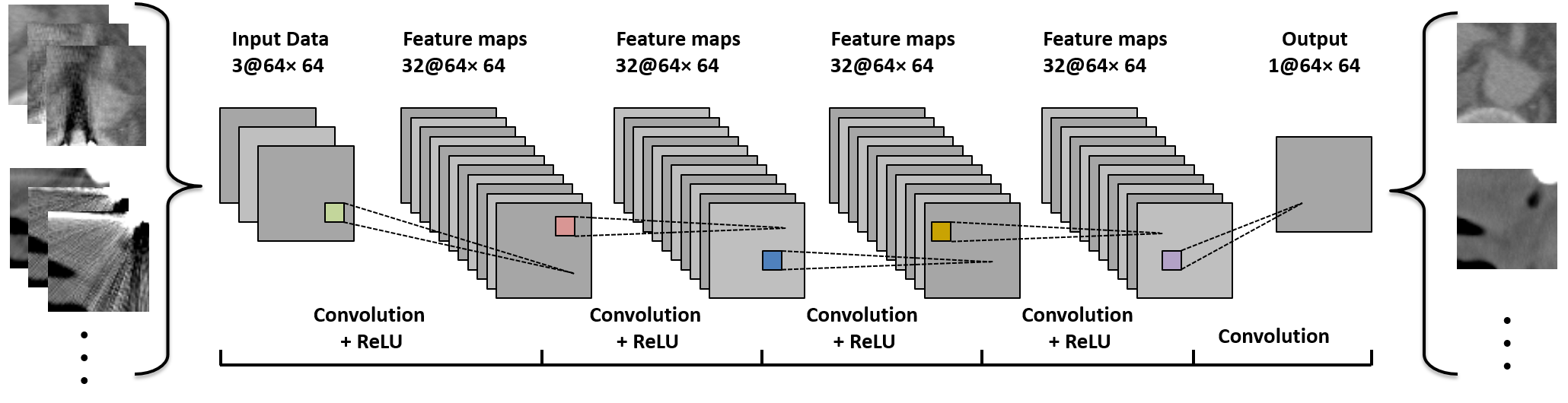}\\
  \caption{Architecture of the convolutional neural network for metal artifact reduction.}
  \label{Fig3_CNN_flowchart}
  \end{center}
\end{figure*}

Fig. \ref{Fig3_CNN_flowchart} depicts the workflow of our CNN, which is comprised of an input layer, an output layer and $ L=5 $ convolutional layers. The ReLU, a nonlinear activation function defined as $\text {ReLU} (x) = \text{max} (0, x)$, is performed after each of the first $L-1$ convolutional layers. In each layer, the output after convolution and ReLU can be formulated as:
\begin{equation}
\label{eq13}
C_l (\mathbf{u}) = \text {ReLU} (\mathbf{W}_l * C_{l-1} (\mathbf{u}) + \mathbf{b}_l)
, l = 1, \dots , L-1 ,
\end{equation}
where $*$ means convolution, $\mathbf{W}_l$ and $\mathbf{b}_l$ denote weights and biases in the $l^{th}$ layer, respectively. We define $C_0 (\mathbf{u}) = \mathbf{u}$. $\mathbf{W}_l$ can be assumed as an $n_l$ convolution kernel with a fixed size of $c_l \times c_l$. $C_l (\mathbf{u})$ generates new feature maps based on the $(l-1) ^ {th}$ layer’s output. For the last layer, feature maps are used to generate an image that is close to the target. Then, we have:
\begin{equation}
\label{eq14}
C_L (\mathbf{u}) = \mathbf{W}_L * C_{L-1} (\mathbf{u}) + \mathbf{b}_L .
\end{equation}
After the construction of the CNN, the parameter set $\Theta = \{ \mathbf{W}_1, \cdots, \mathbf{W}_L, \mathbf{b}_1, \cdots, \mathbf{b}_L \} $ is updated during the training. The estimation of the parameters can be obtained by minimizing the following loss function:
\begin{equation}
\label{eq15}
Loss(\mathbf{U}, \mathbf{V}, \Theta) = \frac {1}{R} \sum _{r = 1} ^R \| C_L(\mathbf{u}_r) - \mathbf{v}_r    \| _F ^2 ,
\end{equation}
where $\mathbf{U} = \{ \mathbf{u}_1, \cdots, \mathbf{u}_R \} $ and $\mathbf{V} = \{ \mathbf{v}_1, \cdots, \mathbf{v}_R \} $ are the input and target datasets, respectively.

\section{CNN-MAR METHOD}
\begin{figure*}
  \begin{center}
  \includegraphics[width = 0.75\textwidth]{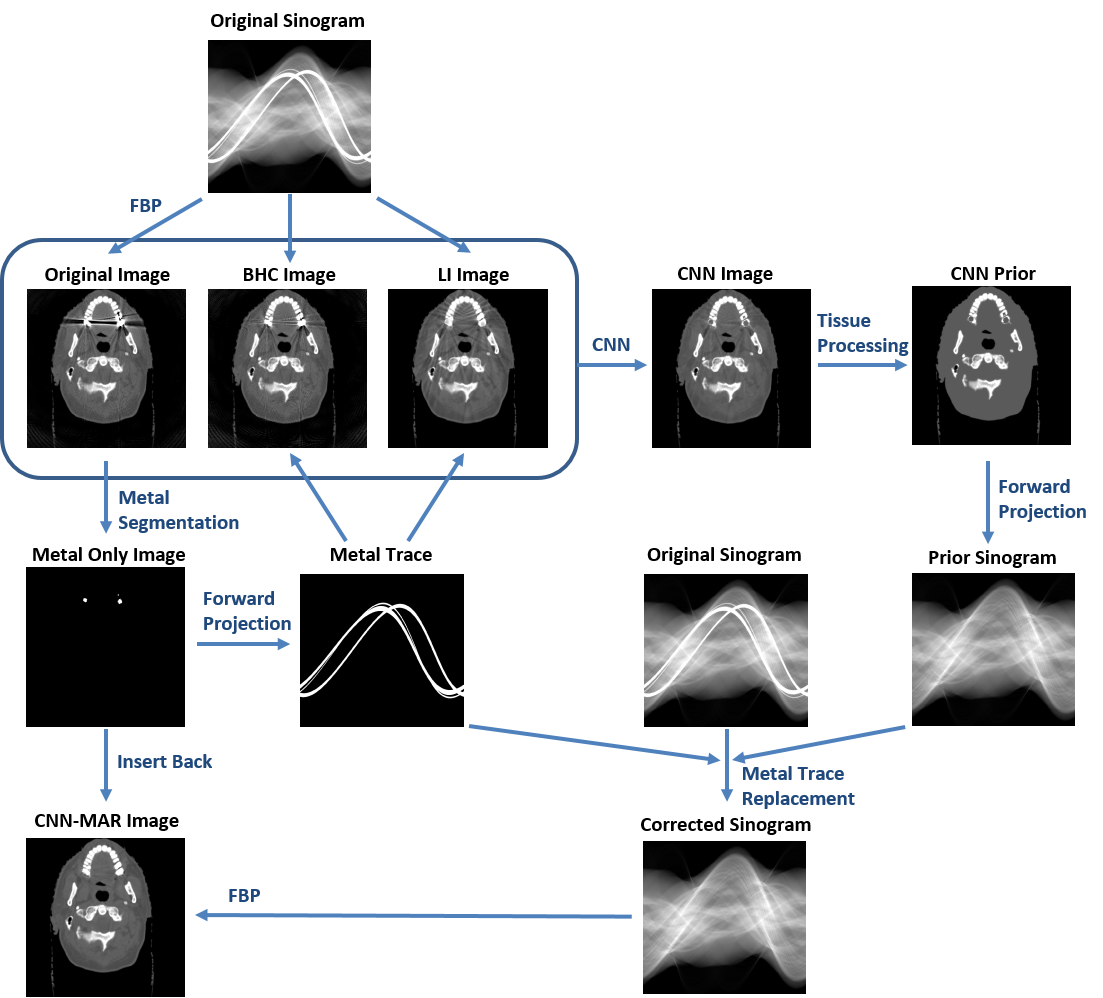}\\
  \caption{Flowchart of the proposed CNN-MAR method.}
  \label{Fig4_CNN_MAR}
  \end{center}
\end{figure*}

Because the proposed MAR approach is based on the CNN, it is referred to as CNN-MAR method. It consists of five steps: (1) metal trace segmentation; (2) artifact reduction with the LI and BHC; (3) artifact reduction with the trained CNN; (4) generation of a CNN prior image using tissue processing; (5) replacement of metal-affected projections with the forward projection of CNN prior, followed by the FBP reconstruction. The workflow of CNN-MAR is summarized in Fig. \ref{Fig4_CNN_MAR}. Steps 1 and 5 are the same as our previous work \cite{HMAR2013}, and step 2 has been described in the above subsection. Hence, we only provide the details for the key steps 3 and 4 as follows.

\subsection{CNN Processing}
\begin{figure}
  \begin{center}
  \includegraphics[width = 0.45\textwidth]{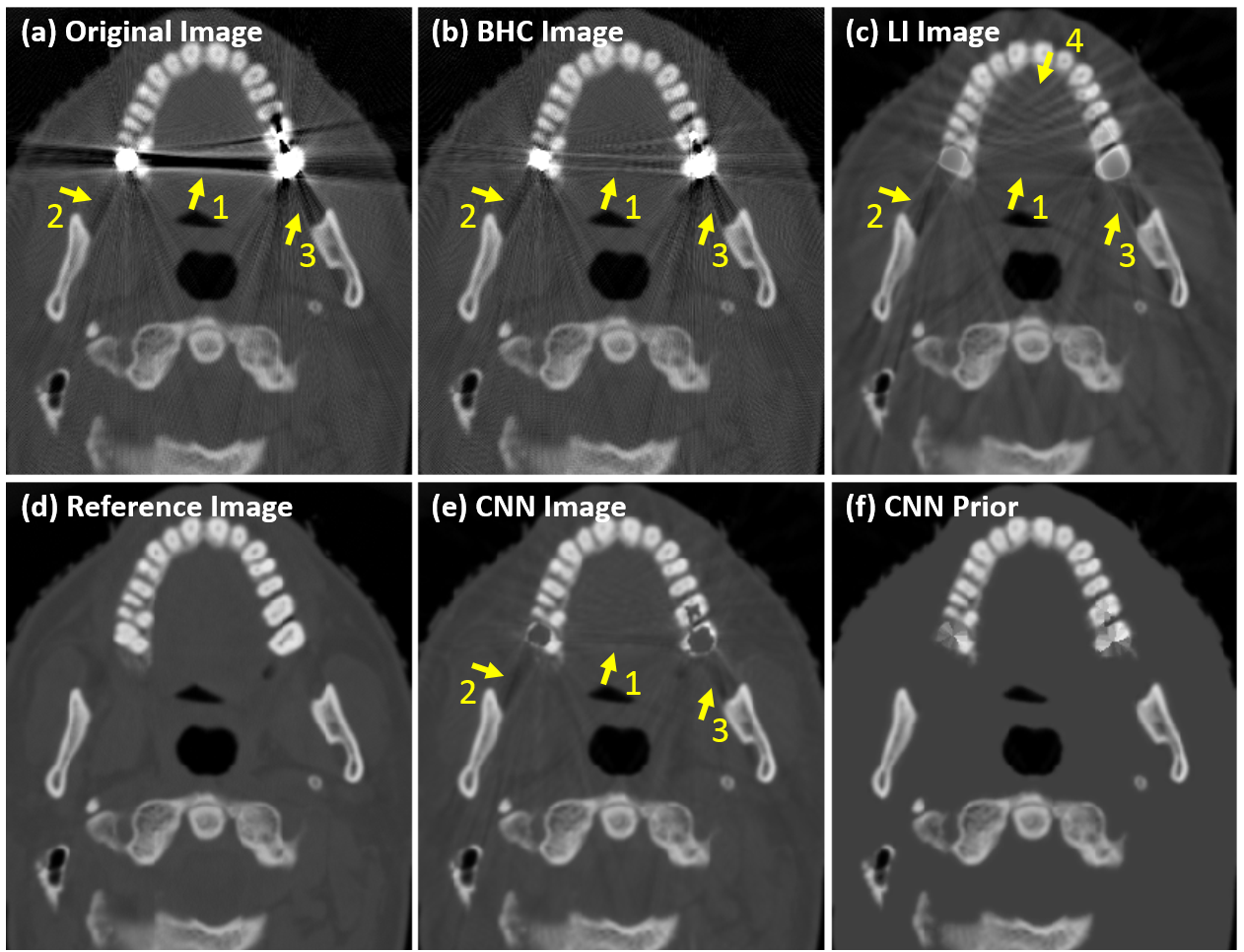}\\
  \caption{Illustration of the CNN image and CNN prior.}
  \label{Fig5_CNNprior}
  \end{center}
\end{figure}

After the BHC and LI corrections, the original uncorrected
image $\mathbf{x}^{art}$, BHC image $\mathbf{x}^{BHC}$ and LI image $\mathbf{x}^{LI}$ are
combined as a three-channel image $\mathbf{x}^{input}$. Hence, the image
after CNN processing is
\begin{equation}
\label{eq16}
\mathbf{x}^{CNN} = C_L(\mathbf{x}^{input}) , 
\end{equation}
where the parameters in $C_L$ have been obtained in advance in the CNN training phase. Fig. \ref{Fig5_CNNprior} shows an example of the CNN inputs and processed CNN image. All the three input images contain obvious artifacts, as indicated by the arrows 1-3. Although the LI alleviates the artifacts indicated by the arrow 1, it introduces new artifacts indicated by the arrow 4. In the CNN image, the artifacts are remarkably suppressed .

\subsection{Tissue Processing}

Although the metal artifacts are significantly reduced after the CNN processing, the remaining artifacts are still considerable. We generate a prior image from the CNN image by the proposed tissue processing approach. Because the water equivalent tissues have similar attenuations and are accounted for a dominate proportion in a patient, we assign these pixels with a uniform value to remove most of the artifacts and obtain a CNN prior image.

By the k-means clustering on the CNN image, two thresholds are automatically determined and the CNN image is segmented into bone, water and air. To avoid wrong clustering in the case of only a few bone pixels, the bone-water threshold is not less than 350 HU. Additionally, to preserve low-attenuated bones, larger regions are segmented with half of the bone-water threshold, and those regions overlapped with the previously obtained bony regions are also assumed as bone and preserved. Then, we obtain a binary image $\mathbf{B}$ for water regions with the target pixels setting to be one and the rest setting to be zero.

Because it may cause discontinuities at boundary and produce fake edges/structures to directly set all water regions with a constant value \cite{Bal2006, HMAR2013}, we introduce an $N=5$ pixel transition between water and other tissues. Based on the binary image $\mathbf{B}$, we introduce a distance image $\mathbf{D}$, where the pixel value is set to be the distance between this pixel and its nearest zero pixel if the distance is not greater than $N$, and is set to be $N$ if it is greater than $N$. Hence, in the image $\mathbf{D}$, most of the water pixels are with the value $N$, and there is an $N$ pixel transition region, while the other tissues are still zeros. We compute the weighted average of water pixel values:
\begin{equation}
\label{eq17}
\bar x ^{CNN, w} = \frac{\sum _i {D_i x^{CNN}_i}}{\sum_i {D_i } } ,
\end{equation}
Thus, the prior image is obtained:
\begin{equation}
\label{eq18}
x^{prior}_i = \frac {D_i}{N} \bar x ^{CNN, w} + (1 - \frac {D_i}{N}) x^{CNN}_i .
\end{equation}
Finally, to avoid the potential discontinuities at boundaries of metals, the metal pixels are replaced with their nearest pixel values.

\begin{figure}
  \begin{center}
  \includegraphics[width = 0.45\textwidth]{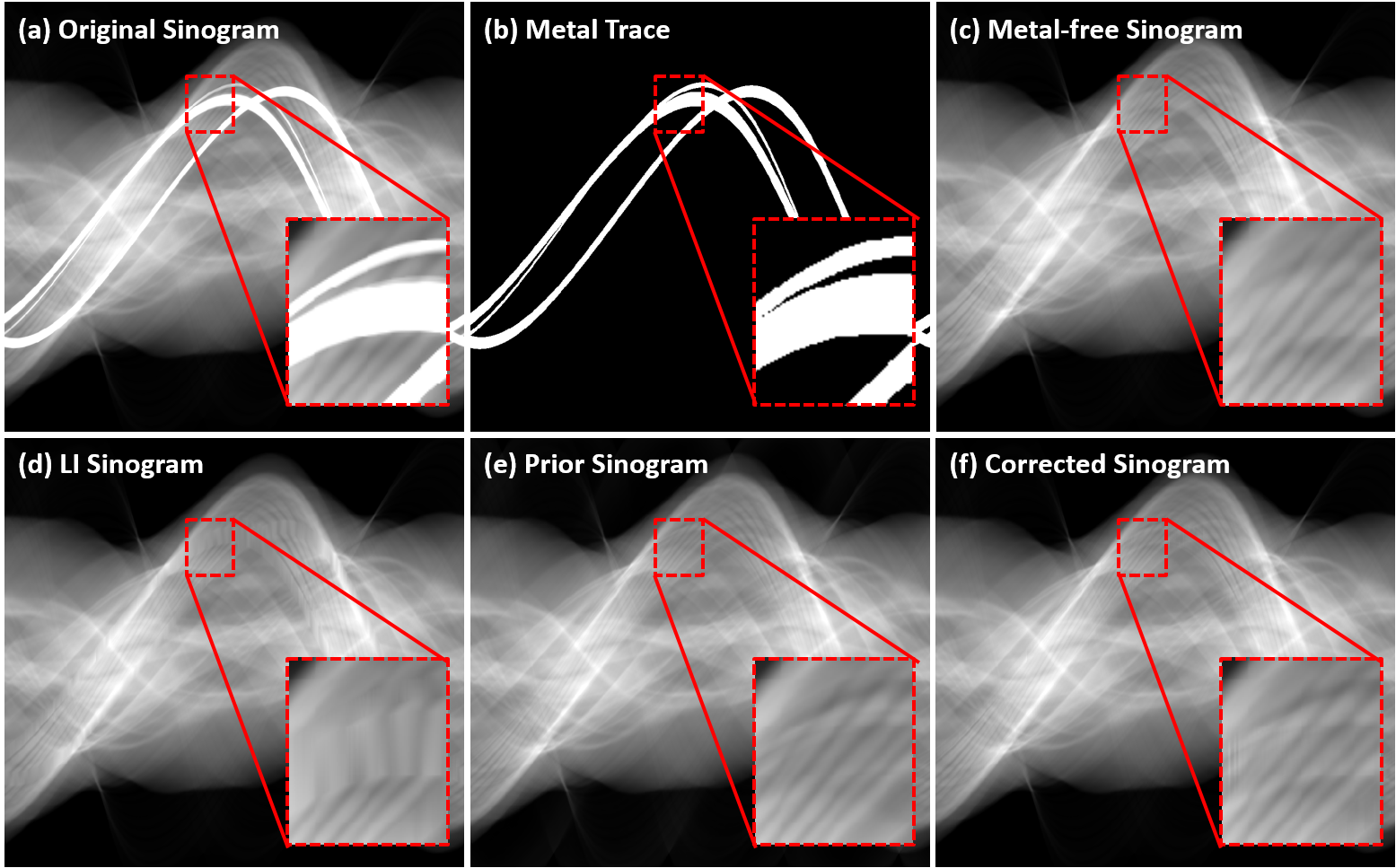}\\
  \caption{Comparison of sinogram completion. An ROI is enlarged and displayed with a narrower window.}
  \label{Fig6_Projection_Compare}
  \end{center}
\end{figure}

Fig. \ref{Fig5_CNNprior}(f) shows an example of the CNN prior image after the tissue processing. It is clear that the regions of water equivalent tissue are flat and the artifacts are removed. Simultaneously, the bony structures are persevered very well. The CNN prior is beneficial for the projection interpolation. As shown in Fig. \ref{Fig6_Projection_Compare}, the LI is a poor estimation of the missing projections. With the help of forward projection of the CNN prior, the surrogate sinogram is extremely close to the ideal one.

\section{Experiments}
\subsection{Creating a Metal Artifact Database}
74 metal-free CT images and 15 metal shapes are collected. Various metal implants are simulated, such as dental fillings, spine fixation screws, hip prostheses, coiling, wires, \emph{etc}. The metal materials include titanium, iron, copper and gold. We carefully adjust the sizes, angles, positions and inserted metal materials so that the simulations are close to clinical cases. In this work, a database is created with 100 cases.

To segment water and bone from a benchmark image, thresholds $T_1$ and $T_2$ are set to linear attenuation coefficients corresponding to 100 HU and 1500 HU, respectively. Mass attenuation coefficients of water, bone and metals are obtained for the XCOM database \cite{xcom}. To simulate metal-free and metal-inserted data, an equi-angular fan-beam geometry is assumed. A 120 kVp x-ray source is simulated and each detector bin is expected to receive $2 \times 10^7$ photons in the case of blank scan \cite{Tang2008}. There are 984 projection views over a rotation and 920 detector bins in a row. The distance between the x-ray source and the rotation center is 59.5 cm. The metal-free and metal-inserted images are reconstructed by the FBP from simulated sinograms and each image consists of $512 \times 512$ pixels.

\subsection{CNN Training} 
In Fig.\ref{Fig3_CNN_flowchart}, the convolutional kernel is $3 \times 3$ in each layer. Therefore, the convolutional weights are $3 \times 3 \times 3 \times 32$ in the first layer, $3 \times 3 \times 32 \times 32$ in the second to the fourth layers and $3 \times 3 \times 32 \times 1$ in the last layer. We set the padding to 1 in each layer so that the size of the output image is the same as the input.

To train the CNN, images are selected from the database to generate the training data. 10,000 patch samples with the size of $64 \times 64$ are extracted from the selected images. Because the spatial distribution of metal artifacts in an image is not uniform, we design a specific strategy to select training patches. A major proportion of the total training data are those patches with strongest artifacts in each corrected image, and the rest patches are randomly selected. The trained neural networks are very similar with different proportions between 50\% to 80\%. The obtained training data are randomly divided into two groups. 80\% of the data is used for training and the rest is for validation during the CNN training. The CNN is implemented in Matlab with the MatConvNet toolbox \cite{Matconvnet, MatconvNetWebsite}. A GeForce GTX 970 GPU is used for acceleration. The training code runs about 25.5 hours and stops after 2000 iterations.

\subsection{Numerical Simulation}
Three typical metal artifacts cases are selected from the database to evaluate the usefulness of the proposed method. They are: case 1, two hip prostheses; case 2, two fixation screws and a round metal inserted in bone; case 3, several dental fillings. These cases are not used in the CNN training.

The proposed method is compared to the BHC, LI and a famous prior image based method NMAR \cite{Meyer2010}. In the NMAR, a prior image is generated from an original image in the case of smaller metal objects of medium density and from an LI image in the case of strong artifacts. For a comprehensive comparison, we generate prior images from both of the original and LI images for the NMAR, which are referred as to NMAR1 and NMAR2 in this paper, respectively. For a quantitative evaluation, we use the metal-free images as references to compute the root mean square error (RMSE) and the structural similarity (SSIM) index \cite{SSIM2004}.

\subsection{Real Data}
The effectiveness of the proposed method is also validated over a clinical data. A patient with a surgical clip is scanned on a Siemens SOMATOM Sensation 16 CT scanner with 120 kVp and 496 mAs using the helical scanning geometry \cite{Yu2007}. The measurement was acquired with 1160 projection views over a rotation and 672 detector bins in a row. The FOV is 25 cm in radius and distance from the x-ray source to the rotation center is 57 cm.

\section{Results}

\subsection{Numerical Simulation}

\begin{figure}
  \begin{center}
  \includegraphics[width = 0.4\textwidth]{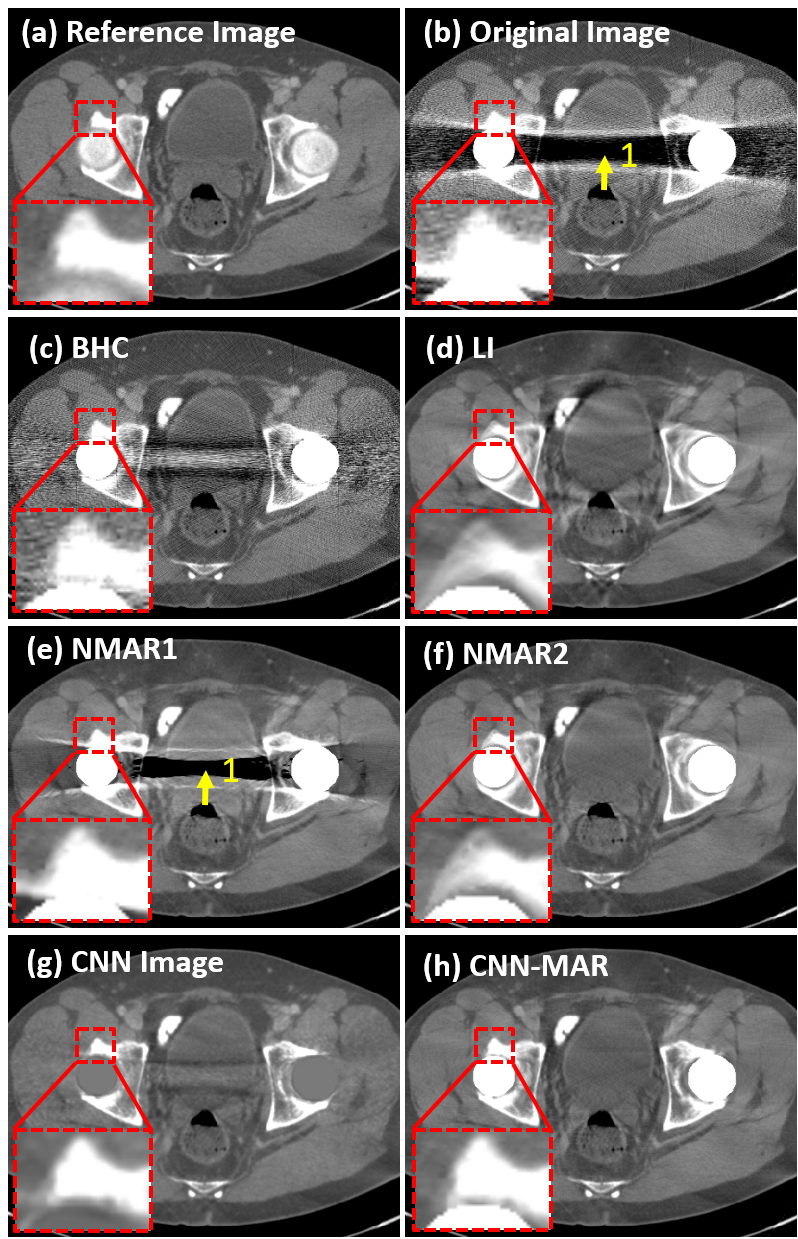}\\
  \caption{Case 1: bilateral hip prostheses. (a) is the reference image, (b) is the original uncorrected image, and (c)-(h) are the corrected results by the BHC, LI, NMAR1, NMAR2, CNN and CNN-MAR, respectively. The ROI highlighted by the small square is magnified. The display window is [-400 400] HU. }
  \label{Fig7_hip_dual}
  \end{center}
\end{figure}

\begin{figure}
  \begin{center}
  \includegraphics[width = 0.45\textwidth]{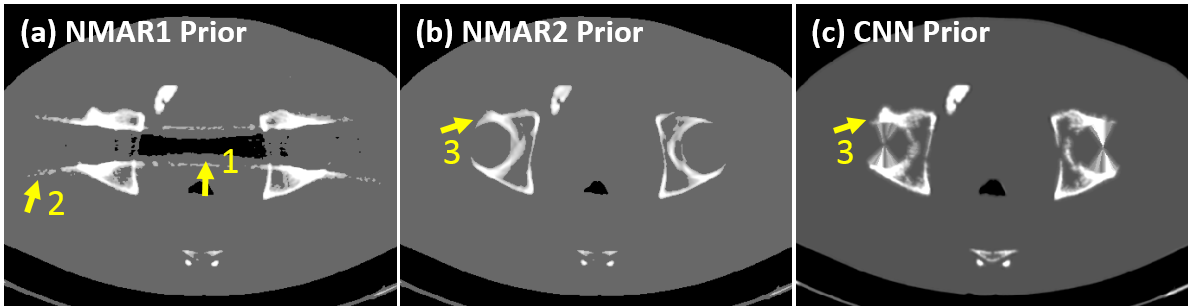}\\
  \caption{The prior images for NMAR1, NMAR2 and CNN-MAR in Fig. \ref{Fig7_hip_dual}.}
  \label{Fig8_hip_prior}
  \end{center}
\end{figure}

Fig. \ref{Fig7_hip_dual} shows the reference, uncorrected and corrected images of the bilateral hip prostheses case. The corresponding prior images for the NMAR1, NMAR2 and CNN-MAR are given in Fig. \ref{Fig8_hip_prior}. A severe dark strip presents between two hip prostheses in the original image as indicated by the arrow ``1''. Although the BHC alleviates these artifacts to some extent, the remaining artifacts are still remarkable. The NMAR1 corrected image also contains strong dark strip in the same location, which is due to its poor prior image. The NMAR method adopts a simple thresholding to segment air, water equivalent tissue, and bone after the image is smoothed with a Gaussian filter \cite{Meyer2010}. Then, air and water regions are set to -1000 HU and 0 HU, respectively. Because of the severe artifacts in the original image, several regions are segmented as wrong tissue types. The NMAR1 prior presents false structures as indicated by the arrows ``1'' and ``2'' in Fig. \ref{Fig8_hip_prior}(a). The false structural information is propagated to the NMAR1 corrected image. The LI corrected image has moderate artifacts compared to the aforementioned methods. However, the bony structures near the metals, as highlighted in the magnified ROI, are blurred and distorted. This is due to the significant information loss near a large metal. As a result, the NMAR2 prior does not suffer from the wrong segmentation but an inaccurate bony structure as indicated by the arrow ``3'' in Fig. \ref{Fig8_hip_prior}(b). Hence, the NMAR2 corrected images reduce artifacts well and introduce wrong bony structures. By comparison, the CNN image captures tissue structures faithfully from the original, BHC and LI images, and avoids most of the artifacts. Due to the excellent image quality of the CNN image, a good CNN prior is generated, followed by a CNN-MAR image with superior image quality. It is clearly seen from Fig. \ref{Fig7_hip_dual}(h) that the artifacts are almost removed completely and the tissue features in the vicinity of metals are faithfully preserved.

\begin{figure}
  \begin{center}
  \includegraphics[width = 0.45\textwidth]{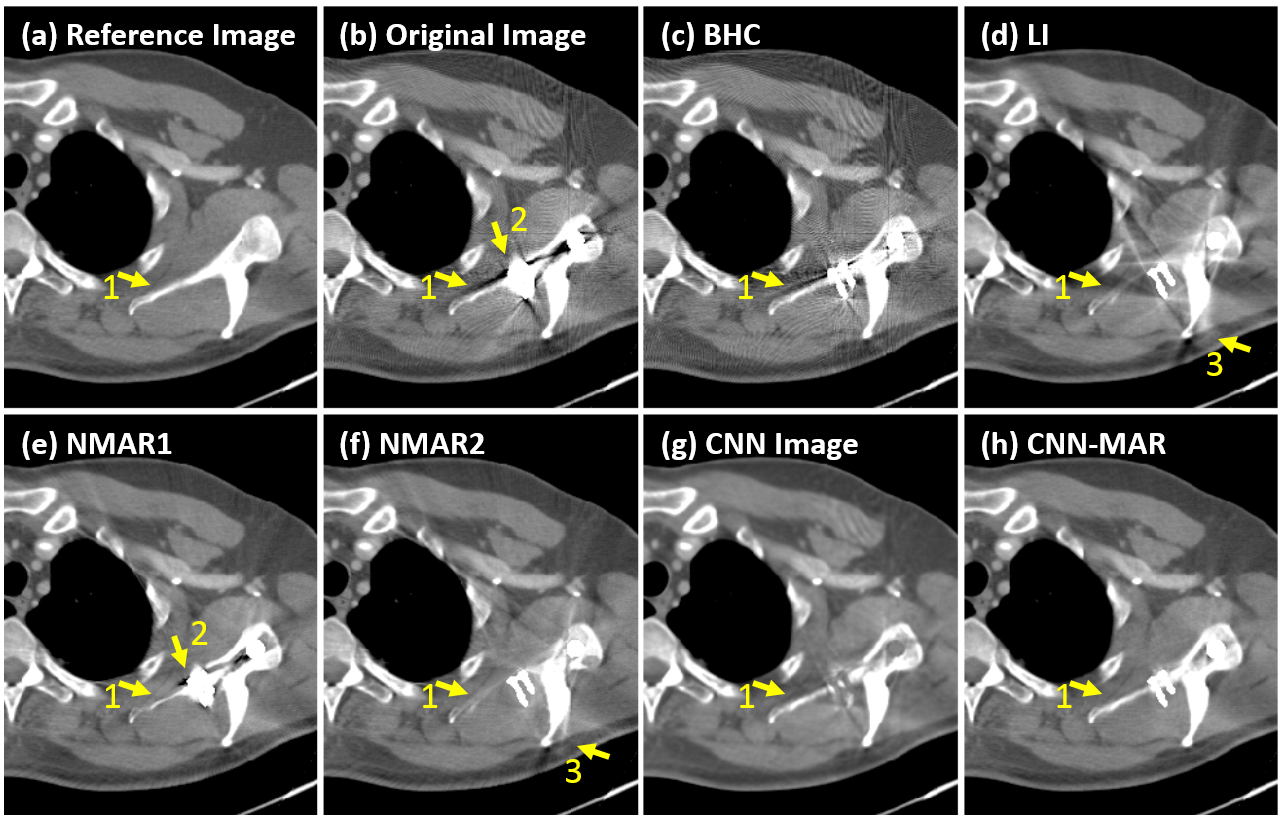}\\
  \caption{Same as Fig. \ref{Fig7_hip_dual} but for case 2: two fixation screws and a metal inserted in the shoulder blade. The display window is [-360 310] HU. }
  \label{Fig9_fixation_metal}
  \end{center}
\end{figure}

Fig. \ref{Fig9_fixation_metal} presents the case 2, where two fixation screws and a metal are inserted in the shoulder blade. The metal artifacts in the original image are moderate, and the BHC is able to remove the bright artifacts (arrow ``2'') around the metals and recovered some bony structures. On the contrary, the LI introduces many new artifacts, and most of the bony structures near the metals are lost as indicated by the arrow ``1''. Both the NMAR1 and NMAR2 are not able to obtain satisfactory results because it can hardly get a good prior image from the original or LI corrected images. The CNN image restores most of the bony features near the metals, and no new artifacts are introduced. Consequently, the CNN-MAR corrected image is very close to the reference.

\begin{figure}
  \begin{center}
  \includegraphics[width = 0.4\textwidth]{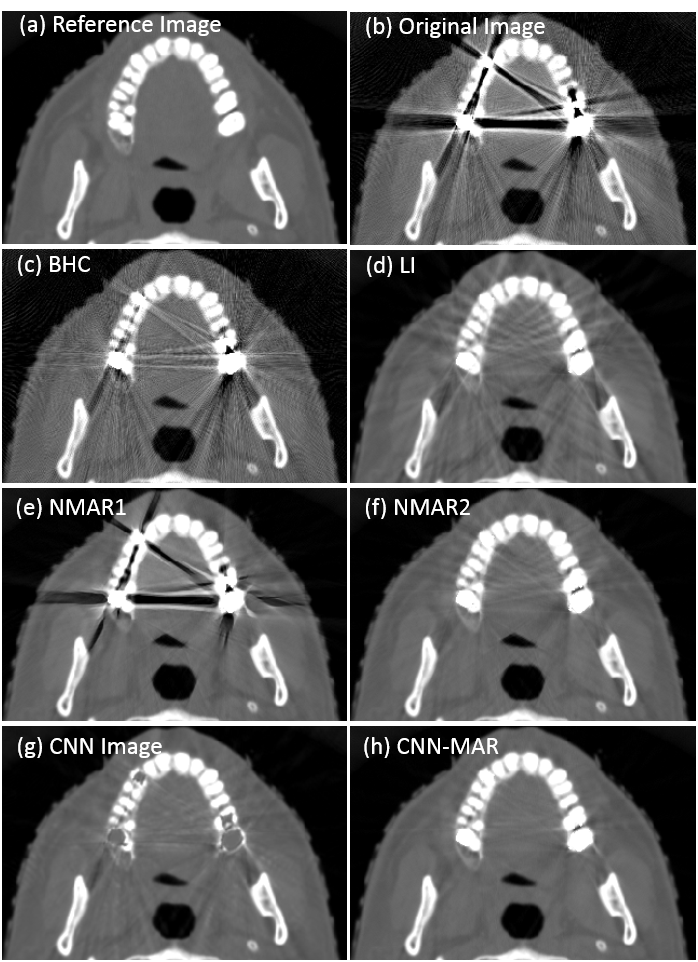}\\
  \caption{Same as Fig. \ref{Fig7_hip_dual} but for case 3: four dental fillings. The display window is [-1000 1400] HU. }
  \label{Fig10_dental_fillings}
  \end{center}
\end{figure}

Fig. \ref{Fig10_dental_fillings} shows the dental images with multiple dental fillings. The original, BHC, LI and NMAR1 images suffer from severe artifacts, and the NMAR2 has less artifacts. Although none of Figs. \ref{Fig10_dental_fillings}(b)-\ref{Fig10_dental_fillings}(d) has a good image quality, the CNN demonstrates an outstanding capacity to preserve the tissue features and avoid most of the strong artifacts simultaneously. Consistent with the previous cases, the CNN-MAR achieves the best image quality.

Table I lists the RMSEs of the original and corrected images with respect to the reference images, where the metallic pixels are excluded. Because the noise also contributes to the RMSE, the artifact induced error is slightly smaller than the values listed in the table. The BHC, LI and NMAR1 have overall large error. In comparison, the NMAR2 achieves a higher accuracy. The CNN images have comparable accuracy to the NMAR2, and the CNN-MAR achieves the smallest RMSEs for all these three cases.

Because the SSIM measures the structural similarity between two images, it is good to evaluate the strength of artifacts \cite{SSIM2004}. The SSIM index lies between 0 and 1, and a higher value means better image quality. Table II lists the SSIM of each image in the numerical simulation study. The BHC has comparable SSIM indices to those of the uncorrected images. The other five MAR methods increase the SSIM  significantly. For the LI, NMAR1, NMAR2 and CNN, their ranks are case-dependent. Generally speaking, the NMAR2 and CNN have better image quality. By comparison, the CNN-MAR has the highest SSIM for the three cases, implying its superior and robust artifact reduction capability.

\begin{table*}[ht]
\renewcommand{\arraystretch}{1.3}
\centering
\label{Table1}
\caption{RMSE of each image in the numerical simulation study. (Unit: HU).}
\begin{tabular}{c c c c c c c c}
\hline
\bfseries  & \bfseries Original & \bfseries BHC & \bfseries LI & \bfseries NMAR1 & \bfseries NMAR2 & \bfseries CNN & \bfseries CNN-MAR\\
\hline 
\bfseries Case 1 & 155.0 & 86.3 & 46.2 & 121.2 & 35.4 & 33.1 & 29.1 \\
\bfseries Case 2 & 71.5  & 44.4 & 54.5 & 50.4  & 41.4 & 31.5 & 22.8 \\
\bfseries Case 3 & 320.3 & 183.5& 107.3 & 234.9 & 82.3& 83.4 & 58.4 \\
\hline
\end{tabular}
\end{table*}

\begin{table*}[ht]
\renewcommand{\arraystretch}{1.3}
\centering
\label{Table2}
\caption{SSIM of each image in the numerical simulation study.}
\begin{tabular}{c c c c c c c c}
\hline
\bfseries  & \bfseries Original & \bfseries BHC & \bfseries LI & \bfseries NMAR1 & \bfseries NMAR2 & \bfseries CNN & \bfseries CNN-MAR\\
\hline 
\bfseries Case 1 & 0.565 & 0.576 & 0.576 & 0.887 & 0.935 & 0.940 & 0.943 \\
\bfseries Case 2 & 0.883 & 0.854 & 0.931 & 0.955 & 0.950 & 0.965 & 0.977 \\
\bfseries Case 3 & 0.522 & 0.536 & 0.886 & 0.833 & 0.942 & 0.932 & 0.967 \\
\hline
\end{tabular}
\end{table*}

\subsection{Clinical Application}

\begin{figure}
  \begin{center}
  \includegraphics[width = 0.4\textwidth]{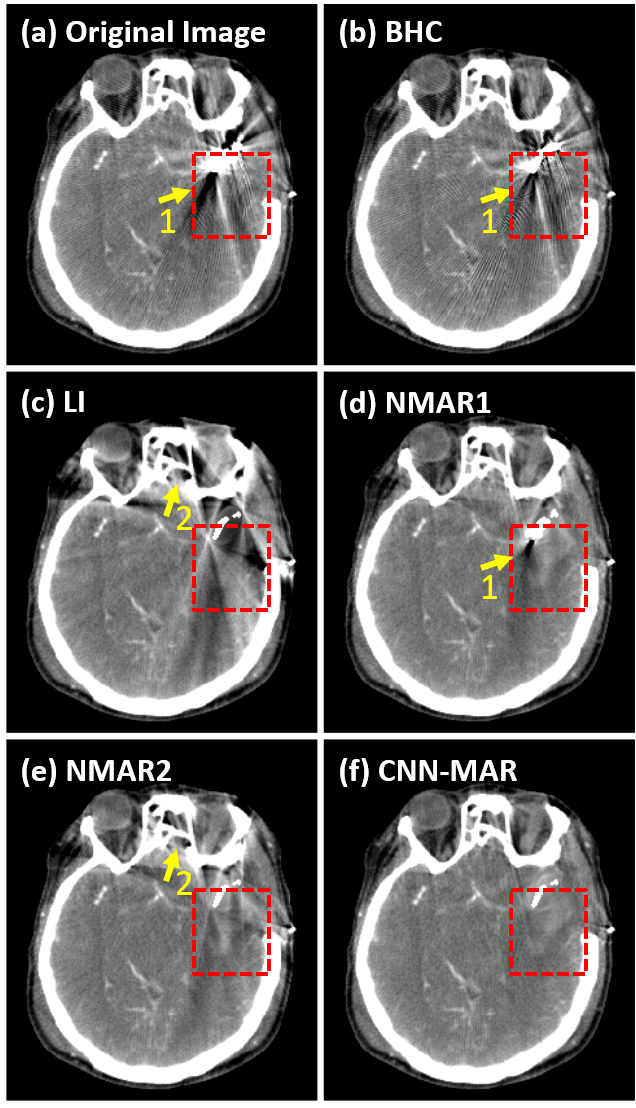}\\
  \caption{The head CT image with a surgical clip. (a) is the original uncorrected image, and (b)-(f) are the corrected results by the BHC, LI, NMAR1, NMAR2 and CNN-MAR, respectively. The display window is [-100 200] HU.}
  \label{Fig11_Real_clip}
  \end{center}
\end{figure}

Fig. \ref{Fig11_Real_clip} shows a patient's head CT image with a surgical clip. The patient is a 59 year-old female with diffused subarachnoid hemorrhage in the basal cisterns and sylvian fissures. The CT angiography demonstrates a left middle cerebral artery aneurysm. She is taken to the operation room and the aneurysm is clipped. She has numerous head CT scans after the surgery for assessment of increased intracranial pressure to rule out rebleeding and hydrocephalus \cite{Yu2007}. The original, BHC and LI images contain too strong artifacts to provide bleeding information in her brain. The NMAR1 and NMAR2 are able to better alleviate artifacts. However, there still exists obvious artifacts in the images as indicated by the arrow ``1'', and bony structures indicated by the arrow ``2'' are distorted. In comparison, the CNN-MAR achieves the best image quality. As highlighted in the rectangular region, there is only one tiny dark streak, and the bright hemorrhage can be observed clearly. The CNN-MAR demonstrates a superior metal artifact reduction and the potential for diagnostic tasks after the clipping surgery.

\subsection{Properties of the Proposed CNN-MAR}

\subsubsection{Effectiveness of the Tissue Processing}

\begin{figure}
  \begin{center}
  \includegraphics[width = 0.45\textwidth]{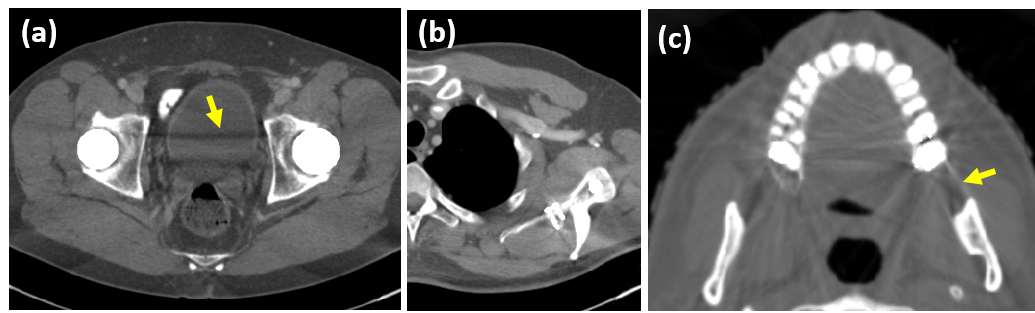}\\
  \caption{Results obtained by directly adopting a CNN image as the prior image without the tissue processing step. (a)-(c) corresponds to the cases 1-3, respectively.}
  \label{Fig12_CNN-Image-MAR2}
  \end{center}
\end{figure}

To study the effectiveness of the tissue processing, we ignore the tissue processing step and directly assume the CNN images as the prior images. The corresponding corrected images are shown in Fig. \ref{Fig12_CNN-Image-MAR2}. Compared to the CNN images in Figs. \ref{Fig7_hip_dual}, \ref{Fig9_fixation_metal} and \ref{Fig10_dental_fillings}, some artifacts can be alleviated by the forward projection. Nevertheless, most of the streaks that are tangent to the metals are preserved as indicated by the arrows in Fig. \ref{Fig12_CNN-Image-MAR2}. By comparison, the tissue processing keeps the major structures and removes the low-contrast features and remaining artifacts. Although the features in the regions of water equivalent tissues are lost after the tissue processing, because the metal-affected projections account for a very small proportion in the sinogram, the missing information is able to be partially recovered from the rest of the unaffected projections. In addition, in the projection replacement step, a projection transition is applied to compensate for the difference between the prior sinogram and the measurements at the boundary of the metal traces \cite{HMAR2013}, which is also beneficial to the information recovery. However, in the presence of large metals, a low-contrast feature in the vicinity of metal may suffer from missing or distortion.

\subsubsection{Selection of Input Images (MAR Methods)}

\begin{figure}
  \begin{center}
  \includegraphics[width = 0.4\textwidth]{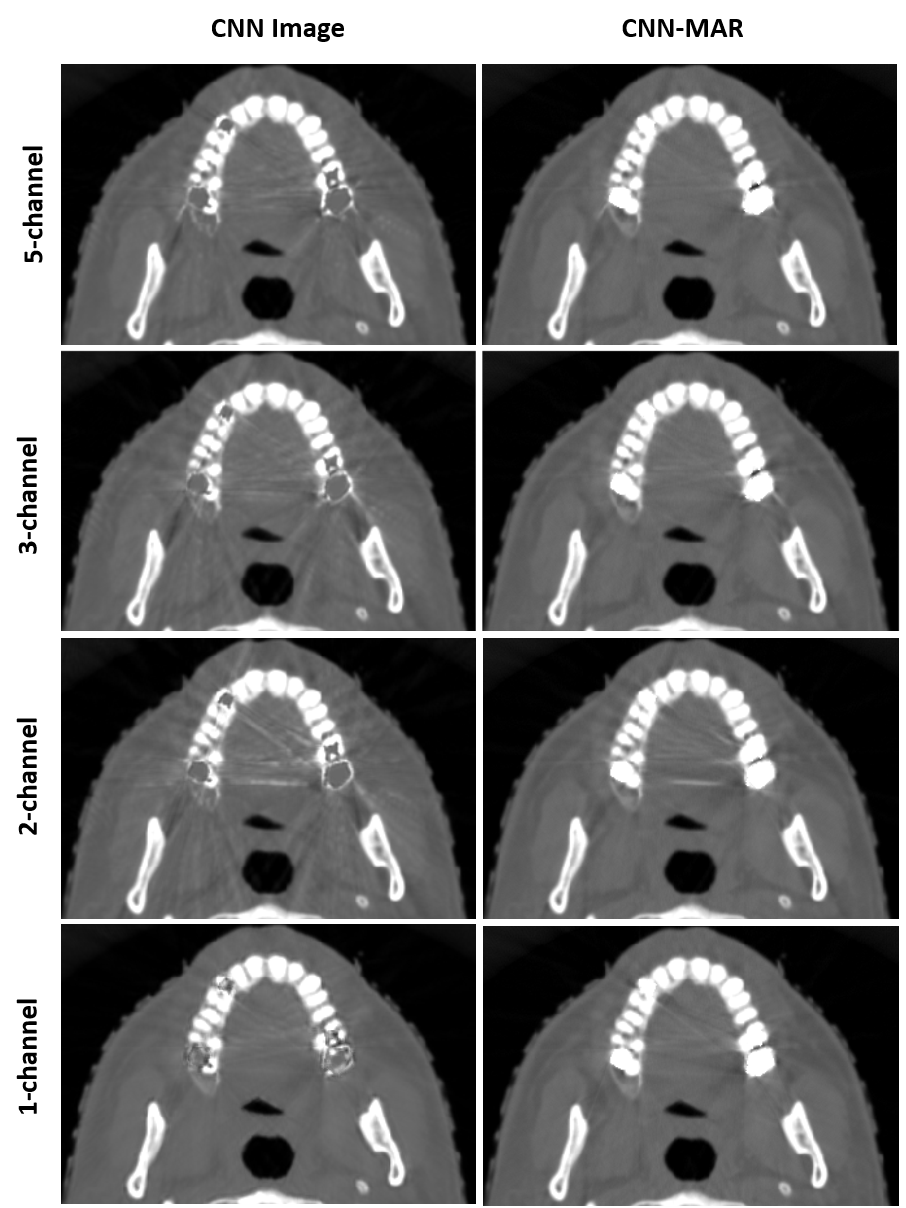}\\
  \caption{CNN and CNN-MAR results based on different channels of input images. Five-channel: original, BHC, LI, NMAR1 and NMAR2 images. Three-channel (default): original, BHC and LI images. Two-channel: original and LI images. One-channel: NMAR2 image.}
  \label{Fig13_MultiMAR_2_5}
  \end{center}
\end{figure}

In this work, the original uncorrected, BHC and LI images are adopted as the input of CNN. We also compare the results with various input images (MAR methods). Here, we apply the original uncorrected, BHC, LI, NMAR1 and NMAR2 images as a five-channel input image, and adopt the original and LI images as a two-channel input image. In addition, the NMAR2 images is employed as a one-channel input image. Fig. \ref{Fig13_MultiMAR_2_5} shows the results of dental fillings case. When NMAR2 is selected as the single input image, the CNN processing is equivalent to the NMAR-CNN method proposed by Gjesteby \textit{et al.}  \cite{Lars_fully3d_2017,Lars_SPIE_image_2017}. Because the NMAR2 image has less artifacts, the CNN image and CNN-MAR image have better image quality. Regarding the multi-channel input, it can be seen from the top three rows that the performance of artifacts reduction is improved by introducing more input images. Particularly, compared to the cases of two-channel input images, three-channel input images remarkably improve the image quality. Therefore, introducing the NMAR1 and NMAR2 only brings limited benefits. This effect depends on if the newly introduced input images contain new useful information. As the aforementioned, the BHC and LI belong to different MAR strategies, which provide complementary information. Without the BHC, some artifacts are wrongly classified as tissue structures and preserved, as illustrated in the third row of Fig. \ref{Fig13_MultiMAR_2_5}. On the contrary, the NMAR1 and NMAR2 are obtained based on prior images from the original and LI images, respectively. Hence, they provide limited new information. In summary, as an open MAR framework, the performance of CNN-MAR can be further improved in the near future by incorporating various types of MAR algorithms.

\subsubsection{Architecture of the CNN}

\begin{figure}
  \begin{center}
  \includegraphics[width = 0.45\textwidth]{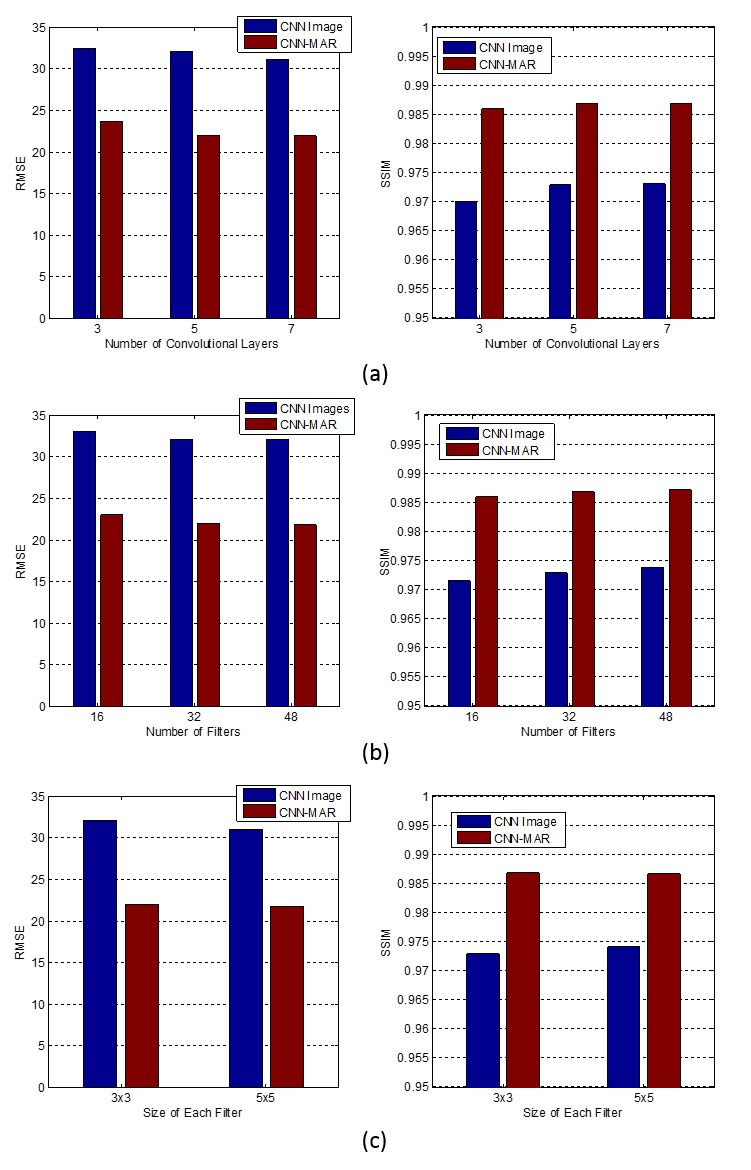}\\
  \caption{Average RMSE and SSIM of CNN images and CNN-MAR images with respect to various CNN architecture parameters: (a) number of convolutional layers, (b) number of filters/features in each layer and (c) the size of each filter. The default CNN has 5 convolutional layers, 32 filters per layer, and each filter is with the size of 3 $\times $ 3. }
  \label{Fig_CNN_para}
  \end{center}
\end{figure}

To study the performance of CNN with respect to different architectures, we adjust the CNN parameters and calculate the average RMSE and SSIM over ten simulated metal artifact cases including the aforementioned three cases. Fig. \ref{Fig_CNN_para} shows the values of RMSE and SSIM using the networks with different number of convolutional layers, number of filters per layer, and the size of each filter. In each subplot, there is only one parameter to be tuned and other parameters are kept as the default ones. The number of neurons in the network increases with the increase of these three parameters, obtaining slightly smaller RMSE and greater SSIM indices. However, because the computational cost rises considerably by using greater parameters, we employ a medium size CNN in this work.

\subsubsection{Training Data}

\begin{figure}
  \begin{center}
  \includegraphics[width = 0.45\textwidth]{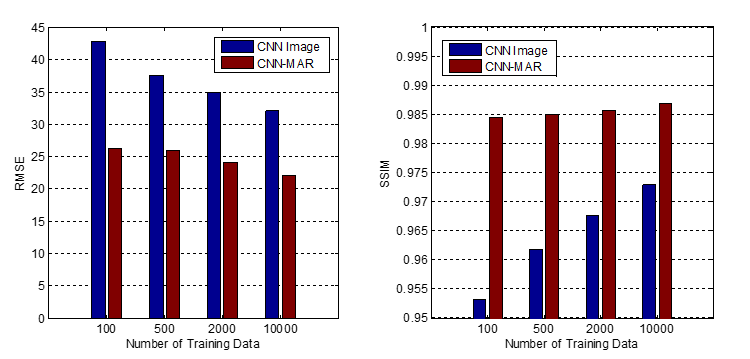}\\
  \caption{Average RMSE and SSIM values using the CNN trained with different data size.}
  \label{Fig_data_bar}
  \end{center}
\end{figure}

We compare the network trained with different numbers of patches. Fig. \ref{Fig_data_bar} presents the average RMSE and SSIM values over ten cases of our results using the network trained with 100, 500, 2000 and 10000 patches. It is clear that the RMSE decreases and the SSIM increases dramatically by applying more training data. This suggests that the performance of the proposed method strongly depends on the size of training data.

\begin{figure}
  \begin{center}
  \includegraphics[width = 0.4\textwidth]{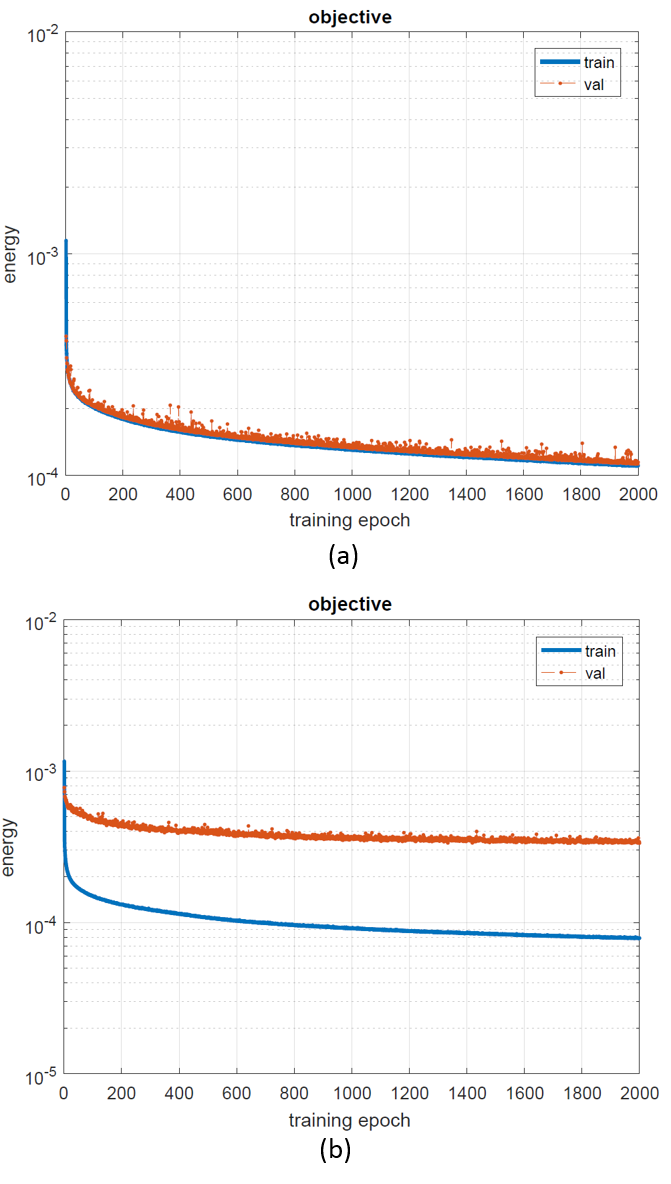}\\
  \caption{The convergence curves of CNN training in terms of energy of loss function versus training epochs. (a) Training data and validation data are selected from the same dataset. (b) Training data and validation data are from different cases in the dataset.}
  \label{Fig14_CNN_curves}
  \end{center}
\end{figure}

We also compare selection strategies for the training data.
The convergence curves of CNN training are presented in Fig. \ref{Fig14_CNN_curves}(a), and the obtained network after 2000 training epochs is used in this work. It can be observed that the energy of the objective function decreases steadily with the increasing training epoch. In Fig. \ref{Fig14_CNN_curves}(a), the training and validation data are selected from a subset of the same dataset, which consists of all types of inserted metals. It is clear that the trained CNN works well on the validation data. In Fig. \ref{Fig14_CNN_curves}(b), the training data and the validation data are selected from the same subset. While the training data is from all types of metals except the multiple dental fillings, and the validation data is from the multiple dental fillings cases. The two separated curves demonstrate an unsatisfactory performance of the obtained CNN on the validation data caused by the difference of the artifact patterns in the two data sets. Hence, it is crucial to include a wider variety of metal artifacts cases as the training data.

\subsubsection{Training Epochs}

\begin{figure}
  \begin{center}
  \includegraphics[width = 0.45\textwidth]{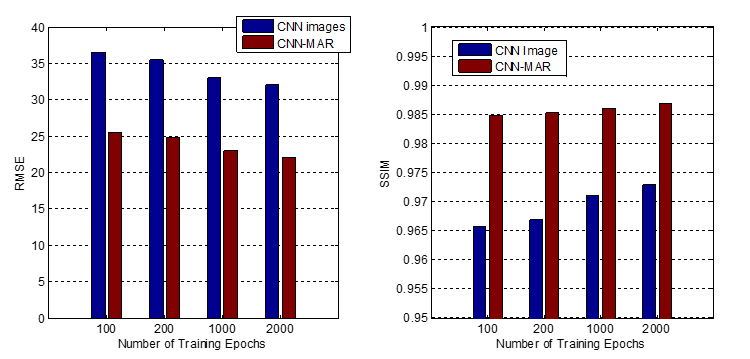}\\
  \caption{Average RMSE and SSIM values using the CNN trained after different epochs.}
  \label{Fig_iter_bar}
  \end{center}
\end{figure}

The proposed method is tested with different training epochs. Fig. \ref{Fig_iter_bar} compares average RMSE and SSIM values over ten cases of the CNN and CNN-MAR images obtained with the network after 100, 200, 1000 and 2000 training epochs. Obviously, by increasing the training epochs, the RMSE of CNN images decreases steadily and the SSIM increases constantly. After the tissue processing, the image quality of CNN-MAR images is remarkably improved. Likewise, the RMSE and SSIM of CNN-MAR images with respective to training epochs follows the same trend to those of CNN images.

\section{Discussion And Conclusion}

From the aforementioned experimental results, it can be seen that the CNN and tissue processing are two mutual beneficial steps. For the CNN step, its strength is to fuse useful information from different sources to avoid strong artifacts. Its drawback is that the CNN can hardly remove all artifacts and mild artifacts typically remain. As to the tissue processing, similar to other prior image based MAR methods, it can remove moderate artifacts and generate a satisfactory prior image. However, in the presence of severe artifacts, the prior image usually suffers from misclassification of tissues. By incorporating the CNN and tissue processing, the CNN training can stop with fewer epochs, and the obtained CNN prior is not affected by tissue misclassification. Their strengths are complementary.

The key factors to ensure outstanding performance of the CNN-MAR are twofold: selection of the appropriate MAR methods and preparation of the training data. The former factor provides sufficient information for the CNN to distinguish tissue structures from the artifacts. The later ensures the generality of the trained CNN by involving as many varieties of metal artifacts cases as possible.

The forward projection of metal identifies which project data is affected. For data correction/estimation based metal artifact reduction methods, including the proposed method, the performance of artifact reduction may be compromised in the case of inaccurate metal segmentation \cite{Stille2013}. Fortunately, a few advanced metal segmentation schemes have been reported \cite{Yu2007,Hegazy2016}, which can be directly applied to the proposed method. Moreover, the deep learning strategy has been widely used for image segmentation \cite{ResNet2016}. Study of applying the neural network for the robust metal segmentation is planned for our future work. 

Although the proposed CNN-MAR in this paper works on 2D image slices, it can be directly extended to 3D volumetric images. Along the new dimension, due to different spatial distribution patterns of tissue structures and artifacts, the 3D version may achieve superior performance. Meanwhile, 3D data will require more training time.

In conclusion, we have proposed a convolutional neural network based metal artifact reduction (CNN-MAR) framework. It is an open artifact reduction framework that is able to distinguish tissue structures from artifacts and fuse the meaningful information to yield a CNN image. By applying the designed tissue processing technique, a good prior is generated to further suppress artifacts. Both numerical simulations and clinical application have demonstrated that the CNN-MAR can significantly reduce metal artifacts and restore fine structures near the metals to a large extent. In the future, we will increase the training data and involve more MAR methods in the CNN-MAR framework to improve its capability. From a broader aspect, the proposed framework has a great potential for other artifacts reduction problems in the biomedical imaging and industrial applications.



%

%
%
%
%
\section*{Acknowledgment}
The authors are grateful to Dr. Ying Chu at Shenzhen University for deep discussions and constructive suggestions. The authors also appreciate Mr. Robert D. MacDougall's proofreading. Part of the benchmark images in our database are obtained from ``the AAPM 2016 Low-dose CT Grand Challenge''.

\ifCLASSOPTIONcaptionsoff
  \newpage
\fi

\bibliographystyle{IEEEtran}

\bibliography{CNNMAR}

\end{document}